\documentclass[aps,pra,preprint,superscriptaddress]{revtex4-1}
\usepackage{amssymb,amsmath}
\usepackage{latexsym,bm}
\usepackage{graphicx}
\usepackage{rotating}
\usepackage{rotating}
\usepackage[dvipsnames]{xcolor}
\usepackage{CJKutf8}
\usepackage{dcolumn}

\DeclareGraphicsExtensions{.jpg}

\begin{document}

\title{Role of hyperfine interaction in Land\'{e} $g$-factors of $^3\!P^o_0$ clock states}
\author{Tingxian Zhang}
\affiliation{State Key Laboratory of Magnetic Resonance and Atomic and Molecular Physics, Wuhan Institute of Physics and Mathematics, Chinese Academy of Sciences, Wuhan 430071, P. R. China}
\affiliation{Institute of Applied Physics and Computational Mathematics, Beijing 100088, P. R. China}
\affiliation{University of Chinese Academy of Sciences, Beijing 100049, P. R. China}
\author{Benquan Lu}
\affiliation{National Time Service Center, Xi'an 710600, P. R. China}
\affiliation{Institute of Applied Physics and Computational Mathematics, Beijing 100088, P. R. China}
\affiliation{University of Chinese Academy of Sciences, Beijing 100049, P. R. China}
\author{Jiguang Li}
\email{li\_jiguang@iapcm.ac.cn}
\affiliation{Institute of Applied Physics and Computational Mathematics, Beijing 100088, P. R. China}
\author{Chengbin Li}
\affiliation{State Key Laboratory of Magnetic Resonance and Atomic and Molecular Physics, Wuhan Institute of Physics and Mathematics, Chinese Academy of Sciences, Wuhan 430071, P. R. China}
\author{Hong Chang}
\affiliation{National Time Service Center, Xi'an 710600, P. R. China}
\affiliation{University of Chinese Academy of Sciences, Beijing 100049, P. R. China}
\author{Tingyun Shi}
\affiliation{State Key Laboratory of Magnetic Resonance and Atomic and Molecular Physics, Wuhan Institute of Physics and Mathematics, Chinese Academy of Sciences, Wuhan 430071, P. R. China}
\author{Zehuang Lu}
\affiliation{MOE Key Laboratory of Fundamental Physical Quantities Measurement, School of Physics, Huazhong University of Science and Technology, Wuhan 430074, P. R. China}

\date{\today}

\begin{abstract}
In the weak-magnetic-field approximation, we derived a general expression of hyperfine-induced Land\'{e} $g$-factors. By using this formula and the multi-configuration Dirac-Hartree-Fock theory, the $g$-factors were calculated for the $3s3p~^3\!P^o_0$ clock state in $^{27}$Al$^+$ and $5s5p~^3\!P^o_0$ in $^{87}$Sr. 
The present results, $\delta g^{(1)}_{\rm hfs}(^3\!P^o_0) = -1.183(6) \times 10^{-3}$ for $^{27}$Al$^+$ and $\delta g^{(1)}_{\rm hfs}(^3\!P^o_0) = 7.78(30) \times 10^{-5}$ for $^{87}$Sr agree with experimental values very well. 
Our theory is also useful to predict hyperfine-induced Land\'e $g$-factors for other atomic systems.

\end{abstract}

\maketitle

\section{Introduction}
High precision has been achieved at the level of $10^{-18}$ for the $^{27}$Al$^+$ ion clock~\cite{Al-clock2010, Al-clock2019} and the $^{87}$Sr optical lattice clock~\cite{Bloom2014, Sr-clock2015}, benefiting from the $J=0$ electronic angular momentum of the $^3\!P^o_0$ and $^1\!S_0$ states involved in the clock transitions for both systems. Hyperfine interaction caused by the nonzero nuclear spins of $^{27}$Al and $^{87}$Sr nuclei, 
however, destroys the spatial symmetry of electronic states, and thus leads to a mix between the $^3\!P^o_{0}$ clock state and other states with same parity but different angular momenta. Consequently, the ``hyperfine-induced" corrections should be evaluated to the frequency shifts of clock transitions~\cite{Boyd2007, Beloy2009, Porsev2017, Derevianko2016}.

The external magnetic field strength is one of essential factors when assessing uncertainties of the clock-transition frequency. Land\'{e} $g$-factor plays a key role in diagnosis of the magnetic field strength. For example, Brewer \textit{et al.} estimated the average strength of the magnetic field in the $^{27}$Al$^+$ ion clock with assistance of their measured differential $g$-factor of the $3s3p~^3\!P^o_0~-~3s^2~^1\!S_0$ clock transition~\cite{Brewer2019}. As mentioned above, the nonzero $g$-factor for the $^3\!P^o_{0}$ clock state arises from hyperfine interaction. This motivated us to study hyperfine-induced Land\'e $g$-factors of clock states.

For the $^{27}$Al$^+$ ion, Rosenband \textit {et al.} measured $g$-factors of the $3s^2~^1\!S_0$ ground state and $3s3p~^3\!P^o_0$ clock state using the quantum logic spectroscopy~\cite{Rosenban2007}. These results are in good agreement with an existing theoretical evaluation by Itano \textit{et al.}~\cite{Itano2007}. The theoretical value was obtained with the multi-configuration Dirac-Hartree-Fock (MCDHF) method, but the computational models were not presented.

Takamoto and Katori reported the first-order Zeeman shift of the $5s5p~^3\!P^o_0~-~5s^2~^1\!S_0$ clock transition for $^{87}$Sr~\cite{Takamoto2003}, from which the differential $g$-factor can be determined for this transition. Boyd {\it {et al.}} studied the effect of hyperfine interaction on the Land\'e $g$-factor of the $^3\!P^o_0$ clock state for $^{87}$Sr~\cite{Boyd2007}, and the corresponding theoretical evaluation was based on the standard Breit-Wills (BW) and modified Breit-Wills (MBW) theory. As discussed in their work, the inability of the BW and MBW theory to simultaneously predict the $^1\!P^o$ and $^3\!P^o$ properties seems to suggest that the theory is inadequate for $^{87}$Sr~\cite{Boyd2007}. Recently, Shi \textit{et al.}~\cite{Shi2015} also measured the differential $g$-factor of the clock transition for $^{87}$Sr with higher precision. To the best of our knowledge, there is no \textit{ab-initio} calculation on the hyperfine-induced Land\'e $g$-factor of the $^3\!P^o_0$ clock state in $^{87}$Sr until now.  

In this work, we derived a general expression of hyperfine-induced Land\'{e} $g$-factors under the weak-magnetic-field condition. The weak-magnetic-field condition means that the total angular momentum of the atomic system remains as a good quantum number. Employing the MCDHF method~\cite{Grant2007, Fischer2016}, we also calculated the hyperfine-induced $g$-factors of the $3s3p~^3\!P^o_0$ clock state in $^{27}$Al$^+$ and $5s5p~^3\!P^o_0$ in $^{87}$Sr. Since hyperfine and Zeeman interactions depend on the different radial regions, high-quality atomic state wave functions in the whole range are required for accurate determination of hyperfine-induced Land\'e $g$-factors. Therefore, electron correlations, not only in the valence shell but also related to the core, were taken into account systematically by using the active space approach~\cite{Jacek2009, Li2012}. We also stressed the importance of the relativistic effects on the atomic parameters concerned. The good agreement between our calculated $g$-factors and experimental values verifies our theoretical method and computational models.

\section{Theoretical method}
\subsection{Zeeman effect of hyperfine levels}
For an $N$-electron atom system with the nonzero nuclear spin ($I\neq0$), we consider the Hamiltonian in the form of the sum of the relativistic Dirac-Coulomb-Breit (DCB) Hamiltonian $H_0$ and the hyperfine interaction $H{\rm_{hfs}}$,
\begin{equation}
H=H_0+H{\rm_{hfs}} \,.
\end{equation}
The DCB Hamiltonian $H_0$ is given by
\begin{equation}
H_0=\sum_{i=1}^{N} \Bigl[ c\,\bm{\alpha}_i\cdot\bm{p}_i+(\beta_i-1)c^2+V{\rm_{nuc}}(r_i) \Bigr]+\sum_{i>j}^{N} \Bigl[\frac{1}{r_{ij}} + B_{ij} \Bigr] \,,
\end{equation}
where $c$ is the speed of light in vacuum, ${\bm \alpha}_i$ and $\beta_i$ are the 4$\times$4 Dirac matrices, $V_{\rm nuc}(r_i)$ is the monopole part of the electron-nucleus interaction, and $B_{ij}$ is the Breit interaction in the low-frequency approximation,
\begin{equation}
B_{ij}=-\frac{1}{2r_{ij}} \bigg[\bm{\alpha}_i \cdot \bm{\alpha}_j+\frac{(\bm{\alpha}_i\cdot \bm{r}_{ij})(\bm{\alpha}_j\cdot \bm{r}_{ij})}{r_{ij}^2}\bigg] \,.
\end{equation}
The hyperfine interaction $H{\rm_{hfs}}$ can be represented as
\begin{equation}
H_{\rm hfs}=\sum_{k\geq1}\bm{T}^{(k)}\cdot\bm{M}^{(k)} \,.
\end{equation}
Here, $\bm{T}^{(k)}$ and $\bm{M}^{(k)}$ are the spherical tensor operators of rank $k$ in the electronic and nuclear space, respectively~\cite{Lindgren1983}. $k=1$ stands for the magnetic dipole hyperfine interaction, and $k=2$ the electric quadrupole hyperfine interaction. The tiny contribution from higher-order terms with $k>2$ is neglected in this work. The electronic tensor operators ${\bm T}^{(1)}$ and ${\bm T}^{(2)}$ read
\begin{gather}
\bm{T^{(1)}}=\sum_{n=1}^N \bm{t}^{(1)}(n)= -\sum_{n=1}^N i \alpha \left(\bm{\alpha}_n \cdot {\bf l}_n \bm{C}^{(1)}(n)\right)r_n^{-2} \,, \\
\bm{T^{(2)}}=\sum_{n=1}^N \bm{t}^{(2)}(n)= -\sum_{n=1}^N \bm{C}^{(2)}(n)r_n^{-3} \,.
\end{gather}
In the equations above, $i$ is the imaginary unit, $\alpha$ is the fine-structure constant, $\bm{C}^{(1)}$ and $\bm{C}^{(2)}$ are the spherical tensor operators, and $\bf{l}$ is the orbital angular momentum operator.

Hyperfine interaction leads to coupling between electronic angular momentum $\bm{J}$ and nuclear spin $\bm{I}$ to total angular momentum $\bm{F}$, i.e., $\bm{F}=\bm{I}+\bm{J}$.
The wave functions of the atomic system $|FM_F\rangle$ are expressed as
\begin{equation}
\label{HFSwavefuction}
|FM_F\rangle=\sum _{\Gamma,J}d_{\Gamma,J}|\Upsilon \Gamma IJFM_F\rangle \,,
\end{equation}
and
\begin{equation}
|\Upsilon \Gamma IJFM_F\rangle=\sum _{M_I,M_J}\langle IJM_IM_J|IJFM_F\rangle|\Upsilon IM_{I}\rangle|\Gamma J M_{J}\rangle \,.
\end{equation}
Here, $|\Gamma J M_{J}\rangle$ and $|\Upsilon IM_{I}\rangle$ are wave functions of the electrons and the nucleus in the atom, respectively. According to the first-order perturbation theory, hyperfine-induced mixing coefficients $d^{(1)}_{\Gamma,J}$ are given by
\begin{equation}
d^{(1)}_{\Gamma,J}=\frac{\langle\Upsilon \Gamma^{'} IJ^{'}FM_F|H {\rm_{hfs}}|\Upsilon \Gamma IJFM_F\rangle}{E_{\Upsilon\Gamma IJFM_F}-E_{\Upsilon\Gamma^{'}IJ^{'}FM_F}}\,,
\end{equation}
where the prime stands for the perturbing states. The matrix elements for the magnetic dipole hyperfine interaction are
\begin{equation}
\begin{split}
&\langle\Upsilon \Gamma IJFM_F|\bm{T}^{(1)}\cdot\bm{M}^{(1)}|\Upsilon \Gamma^{'} IJ^{'}FM_F\rangle   \\
=&(-1)^{I+J+F} \begin{Bmatrix}I&J&F\\J^{'}&I&1 \end{Bmatrix}\sqrt {2J+1}\sqrt {2I+1}\langle\Gamma J\|\bm{T}^{(1)}\|\Gamma^{'} J^{'}\rangle\langle\Upsilon I\|\bm{M}^{(1)}\|\Upsilon I \rangle\,,
\end{split}
\end{equation}
and for the electric quadrupole hyperfine interaction
\begin{equation}
\begin{split}
&\langle\Upsilon \Gamma IJFM_F|\bm{T}^{(2)}\cdot\bm{M}^{(2)}|\Upsilon \Gamma^{'} IJ^{'}FM_F\rangle  ~~~~~~~~~~~~~~~~~~~~~\\
=&(-1)^{I+J+F} \begin{Bmatrix}I&J&F\\J^{'}&I&2 \end{Bmatrix}\sqrt {2J+1}\sqrt {2I+1}\langle\Gamma J\|\bm{T}^{(2)}\|\Gamma^{'} J^{'}\rangle\langle\Upsilon I\|\bm{M}^{(2)}\|\Upsilon I \rangle\,.
\end{split}
\end{equation}
The nuclear matrix elements $\langle \Upsilon I||\bm{M}^{(1)}||\Upsilon I\rangle$ and $\langle \Upsilon I||\bm{M}^{(2)}||\Upsilon I\rangle$ are related to nuclear magnetic dipole moment $\mu_I$ and electric quadrupole moment $Q_I$ through~\cite{Parpia1996}
\begin{gather}
\langle \Upsilon I I |M^{(1)}_0| \Upsilon I I \rangle =\mu_I  \,,\\
\langle \Upsilon I I |M^{(2)}_0| \Upsilon I I \rangle =\frac{Q_I}{2} \,.
\end{gather}

The Zeeman interaction between an atom and external magnetic field $\bm{B}$ can be written as~\cite{KTCheng1985, HFSzeeman}
\begin{equation}
H{\rm_m}= - {\bm \mu}^{(1)} \cdot {\bm B} + H{\rm_m^{nuc}} \,.
\end{equation}
The electronic tensor operator ${\bm \mu}^{(1)}$ is given by
\begin{equation}
{\bm \mu}^{(1)} = - \frac{1}{2} \left[ {\bm N}^{(1)} + \Delta {\bm N}^{(1)} \right]
\end{equation}
and
\begin{gather}
\bm{N}^{(1)}=\sum_{j=1}^N\bm{n}^{(1)}(j)=-\sum_{j=1}^N \frac{i}{\alpha} \Big(\bm{\alpha}_j\cdot{ \bf l }_j\bm{C}^{(1)}(j)\Big)r_j \,, \\
\Delta \bm{N}^{(1)}=\sum_{j=1}^N\Delta\bm{n}^{(1)}(j)=\sum_{j=1}^N (g_s-2)\beta_j\bm{\Sigma}_j\,,
\end{gather}
where ${\bm \Sigma}_j$ is the relativistic spin-matrix and $g_s = 2.00232$ the $g$-factor of the electron spin corrected by quantum electrodynamic (QED) effects.

In the weak-magnetic-field approximation, 
the total angular momentum $F$ is still a good quantum number for the atomic system and the energy shift of a given hyperfine level $|FM_F\rangle$ can be calculated by 
\begin{align}
\label{zeeman}
\Delta E &= \frac{1}{2} \langle FM_F|N^{(1)}_0+\Delta N^{(1)}_0|  FM_F\rangle B + \langle H^{\rm nuc}_{\rm m} \rangle \nonumber \\
             &= g_F \mu_B M_F B + \langle H^{\rm nuc}_{\rm m} \rangle \,.
\end{align}
Here, the Bohr magneton $\mu_{\rm B} (= e\hbar/2m_e$) is equal to 1/2 in atomic unit.
Substituting Eq. (\ref{HFSwavefuction}) into the equation above, the Land\'{e} $g$-factor can be written as
\begin{align}
g_F \approx &  \frac{\langle\Upsilon \Gamma IJFM_F|N^{(1)}_0+\Delta N^{(1)}_0| \Upsilon \Gamma IJFM_F\rangle}{M_F} + 2\sum _{\Gamma',J'} d^{(1)}_{\Gamma',J'}\frac{\langle\Upsilon \Gamma IJ FM_F|N^{(1)}_0+\Delta N^{(1)}_0| \Upsilon \Gamma' I J' FM_F\rangle} {M_F} \nonumber \\
=& g_0+\delta g^{(1)}{\rm_{hfs}} \,.
\end{align}
The last term $\delta g^{(1)}{\rm_{hfs}}$ represents a hyperfine-induced Land\'e $g$-factor. The involved Zeeman matrix elements between hyperfine states are given by
\begin{multline}
\langle \Upsilon \Gamma IJFM_F|N^{(1)}_0+\Delta N^{(1)}_0|\Upsilon \Gamma^{'}I J^{'} F M_F \rangle\\
= (-1)^{I+J^{'}+1+F} M_F \sqrt {\frac{2F+1}{F(F+1)}}\begin{Bmatrix}J&F&I\\F&J^{'}&1 \end{Bmatrix} \sqrt{2J+1}\langle \Gamma J||{\bm N}^{(1)}+\Delta \bm{N}^{(1)}|| \Gamma^{'} J^{'}\rangle\,,
\end{multline}
where $J'=J-1,\ J,\ J+1$.

\subsection{Hyperfine-induced Land\'{e} $g$-factor of $nsnp~^3\!P^o_0$ clock states \label{g}}
As mentioned above, the nonzero Land\'{e} $g$-factor for an $nsnp~^3\!P^o_0$ clock state is attributed to the hyperfine interaction. Treating the adjacent $^3\!P^o_1$ and $^1\!P^o_1$ states as only perturbing states, and neglecting others because of their fractional contribution due to large energy intervals,
we have the hyperfine-induced Land\'e $g$-factor 
\begin{align}
\label{G3P0}
\delta g^{(1)}_{\rm hfs}(^3\!P^o_0) \approx& 2 \Bigl[ \frac{\langle ^{3}\!P^o_0~FM_F |N^{(1)}_0+\Delta N^{(1)}_0| ^3\!P^o_1~FM_F \rangle}{M_F} \frac{\langle ^{3}\!P^o_1~FM_F|H_{\rm hfs}| ^3\!P^o_0~FM_F\rangle}{(E_{^3\!P^o_0}-E_{^{3}\!P^o_1})} \nonumber \\
&+ \frac{\langle ^{3}\!P^o_0~FM_F|N^{(1)}_0+\Delta N^{(1)}_0| ^1\!P^o_1~FM_F\rangle}{M_F} \frac{\langle ^{1}\!P^o_1~FM_F|H_{\rm hfs}| ^3\!P^o_0~FM_F\rangle}{(E_{^3\!P^o_0}-E_{^{1}\!P^o_1})}\Bigr].
\end{align}

The effect of hyperfine interaction on the $ns^2~^1S_0$ ground state is negligible with respect to large energy separations from other even-parity excited states. In addition, the nuclear Zeeman shifts $\langle H^{\rm nuc}_{\rm m} \rangle$ cancel out between the lower and upper states. As a result, the differential $g$-factor of the $nsnp~^3\!P^o_0~-~ns^2~^1\!S_0$ clock transition is equivalent to the hyperfine-induced Land\'e $g$-factor of the upper $^3\!P^o_0$ clock state.  

\subsection{MCDHF method}
The MCDHF method is utilized to generate electronic state wave functions (ESFs) $|\Gamma J M_J \rangle$~\cite{Grant2007, Fischer2016}. An electronic state wave function is constructed by configuration state functions (CSFs) $|\gamma J M_J \rangle$ with the same parity $P$, electronic total angular momentum $J$ and its component along the $z$ direction $M_J$,
\begin{equation}
\label{MCDHF}
|\Gamma JM_J \rangle = \sum_{i=1}^{NCSF} c_i  |\gamma _i JM_J \rangle \,.
\end{equation}
Here, $c_i$ is the mixing coefficient, $\gamma_i$ stands for other appropriate quantum number of the CSF. Each CSF is a linear combination of products of one-electron Dirac orbitals. The mixing coefficients and the orbitals are optimized simultaneously in the self-consistent field (SCF) procedure to minimize energies of levels concerned. Once a set of orbitals is obtained, the relativistic configuration interaction (RCI) calculations can be carried out to capture more electron correlations, and to include the Breit interaction and QED corrections. In practice, we employed the GRASP2K~\cite{Jonsson2013} and HFSZEEMAN~\cite{HFSzeeman} packages to perform the calculations.

\section{Calculations and Results}


\subsection{The case of $^{27}$\!Al$^+$~\label{model-Al}}

As a starting point, the Dirac-Hartree-Fock (DHF) calculation was performed. At this stage, the orbitals occupied in the $1s^22s^22p^63s3p$ reference configuration were optimized as spectroscopic orbitals. Following that, we considered in the SCF procedures the correlation between the $3s$ and $3p$ electrons in the valence subshells, and the correlation between these two electrons and those in the $n \le 2$ core shells. The former is referred to as the valence-valence (VV) correlation and the latter as the core-valence (CV) correlation. The VV and CV were accounted for by CSFs generated by single (S) and double (D) replacements of the occupied orbitals with virtual orbitals. A restriction was applied so that only one core orbital can be substituted at a step. The virtual orbitals were augmented layer by layer up to $n=13$, $l=5$, and only the last added virtual orbitals were variable in the SCF calculations. This computational model was labelled as CV. The core-core (CC) electron correlation in the $n=2$ shell, labelled as CC$_2$, was further taken into account in the RCI computation. The CSFs produced by exciting one and two electrons from the $n=2$ shell to all virtual orbitals were added into the CV model. The orbitals obtained with the CV model were fixed in the RCI computations. 

The CSFs in CC$_2$ model capture the main first-order electron correlations~\cite{FroeseFischer1997, Li2012}. In order to achieve satisfactory accuracy for physical quantities concerned, the effect of higher-order electron correlations must be considered. For this purpose, we adopted the MR-SD approach~\cite{JGLi2016}, in which the SD-excitation CSFs from the set of multi-reference (MR) configurations are further included. The MR set was formed by incorporating into the $1s^22s^22p^63s3p$ configuration the dominant configurations in the CC$_2$, that is, $2s^22p^63p3d$, $2s^22p^43s3p4p^2$, and $2s^22p^43s3p4d^2$. 


To demonstrate the electron correlation effects on the $g$-factors for Al$^+$, the calculated $g(^{3,1}\!P^o_1)$ in various computational models are presented in Table \ref{g-Al}. The ``Breit" stands for the final results with inclusion of the Breit interaction, and the uncertainties are shown in the parentheses. It was found that the VV and CV electron correlations make main contribution to $g$-factors of both states. In addition, the effect of the CC correlation in the $n=2$ shell on $g(^{3,1}\!P^o_1)$ is opposite to that from the higher-order correlation. The contribution from the neglected electron correlations was estimated to be about $2 \times 10^{-6}$ for $g(^{3}\!P^o_1)$ and $1.7 \times 10^{-6}$ for $g(^{1}\!P^o_1)$. The QED corrections to the $g$-factors ($\sim 2 \times 10^{-8}$) are negligible. 

The non-relativistic $g$-factors of the $^{3}\!P^o_1$ and $^{1}\!P^o_1$ states are 1.5 and 1, respectively, according to the formula $g_{\rm NR} =1+\frac{J(J+1)-L(L+1)+S(S+1)}{2J(J+1)}$ in the $LS$ coupling scheme. Here, the $L$, $S$, and $J$ correspond to the orbital, spin and total angular momentum of electrons. Comparing this value with the DHF result, we found that the one-electron relativistic effect is 0.073\% for the $g(^{3}\!P^o_1)$ and 0.004\% for the $g(^{1}\!P^o_1)$.

Guggemos \textit{et al.} measured the $g$-factor of hyperfine state $3s3p~^{3}\!P^o_1$ with $F=7/2$ in the $^{27}$Al$^+$ ion by the quantum logic spectroscopy, and reported two results: 0.42884(1)~\cite{Guggemos2017} and 0.428133(2)~\cite{Guggemos2019}. Note that the discrepancy between these two values exceeds the experimental error bars of both cases. Using $g(^3\!P^o_1)$ of the ``Breit'' model in Table~\ref{g-Al}, we obtained $g(^{3}\!P^o_1, F=7/2)=0.4288807$ based on
\begin{equation}
g_{F} =\frac{F(F+1)+J(J+1)-I(I+1)}{2F(F+1)}g_{J} \,.
\end{equation}
Our result is in good agreement with the first experimental value. Furthermore, we evaluated the effect of the hyperfine-induced mixing between $^3\!P^o_1$ and $^3\!P^o_2$ states on $g(^{3}\!P^o_1, F=7/2)$. This correction is about $ - 1.798 \times 10^{-4}$, and the resulting $g$-factor is still closer to the first measurement~\cite{Guggemos2017}.


\begin{table}
\caption{Land\'e $g$-factors of the $3s3p~^{3,1}\!P^o_1$ states in Al$^+$. The number in parentheses stands for uncertainties. \label{g-Al}}
\begin{ruledtabular}
\begin{tabular}{ldldl}
Model           && $g(^{3}\!P^o_1)$      && $g(^{1}\!P^o_1)$   \\
\hline
DHF             && 1.5010941             && 0.9999617          \\
CV              && 1.5010785             && 0.9999687          \\
CC$_{2}$        && 1.5010833             && 0.9999657          \\
MR              && 1.5010818             && 0.9999668          \\
Breit           && 1.5010825(20)         && 0.9999662(17)      \\
NR              && 1.5                   && 1.0                \\
\end{tabular}
\end{ruledtabular}
\end{table}

The calculated hyperfine-induced Land\'e $g$-factors of the $^3\!P^o_0$ state are displayed in Table \ref{data-Al} as a function of computational models. The off-diagonal matrix elements of Zeeman and hyperfine interactions, $\langle ^3\!P^o_0 || -{\bm \mu}^{(1)} || ^{3,1}\!P^o_1 \rangle$ and $\langle ^{3,1}\!P^o_1 \|\bm {T}^{(1)}\|^3\!P^o_0\rangle$, and energy separations between the $3s3p~^{3,1}\!P^o_1$ and $3s3p~^{3}\!P^o_0$ states, $\Delta E(^{3,1}\!P^o_1$ - $^{3}\!P^o_0$), are need to calculate $\delta g^{(1)}_{\rm hfs}(^3\!P^o_0)$ (see Eq. (\ref{G3P0})). Therefore, we also present these results in this table. It should be emphasized that we removed those CSFs not interacting with the reference configurations for computational efficiencies when calculating the hyperfine interaction and Zeeman matrix elements. However, the corrections from these removed CSFs must be considered to the energy separations~\cite{Li2012}. As can be seen, the VV and CV electron correlations make dominant contributions to all of the physical quantities under investigation. The CC and higher-order electron correlation effects, although tiny, are non-negligible. Note that the Breit interaction is also significant to improve the fine-structure splitting between $^3\!P^o_1$ and $^3\!P^o_0$ states. The off-diagonal hyperfine interaction matrix elements obtained with the ``Breit" model are consistent with other theoretical results except for ones by Andersson \textit{et al.}~\cite{Andersson2010}. The discrepancies arise from neglected CV correlation related to the $1s$ electrons and CC correlation in the $n=2$ shell in their calculation. A good agreement was found for the fine-structure splitting $\Delta E(^3\!P^o_1 - ^3\!P^o_0)$, while the energy interval between $^1\!P^o_1$ and $^3\!P^o_0$ deviates from the NIST value~\cite{NIST} by 4\%. For the latter is attributed to so-called $LS$-term dependence of the $3p$ valence orbital~\cite{FroeseFischer1997}. To improve this term separation one would optimize $^3\!P$ and $^1\!P$ terms separately. Nevertheless, the resulting orbital bases for these two terms are non-orthogonal with each other. Furthermore, the off-diagonal Zeeman and hyperfine interaction matrix elements cannot be dealt with by using the standard Racah technique~\footnote{The recent developed biorthonormal transformation method will be a good solution for this problem in the near future~\cite{Olsen1995, Verdebout2010, FroeseFischer2013a, Verdebout2013}}. Fortunately, the contribution from the $3s3p~^1\!P^o_1$ perturbing state and other higher excited states are less than $10^{-8}$. Thus, the less good energy interval between the $^1\!P^o_1$ and $^3\!P^o_0$ states does not impact to the final $g$-factors at present accuracy. Our hyperfine-induced Land\'e $g$-factor, $\delta g^{(1)}_{\rm hfs}(^3\!P^o_0) = 1.183 \times 10^{-3}$ is in excellent agreement with the experimental value for $^{27}$Al$^{+}$. The computational uncertainty comes from the rest of electron correlations, especially related to the innermost $1s$ electrons, which was estimated to be about $6 \times 10^{-6}$.

\begin{turnpage}
\begin{table}
\caption{Matrix elements (in a.u.) of Zeeman and hyperfine interactions, $\langle^3\!P^o_0 ||-{\bm \mu}^{(1)}||^{3,1}\!P^o_1\rangle$ and $\langle^{3,1}\!P^o_1 \|\bm {T}^{(1)}\|^3\!P^o_0\rangle$, energy intervals (in a.u.) $\Delta E(^{3,1}\!P^o_1~-~^{3}\!P^o_0$), and $\delta g^{(1)}_{\rm hfs}(^3\!P^o_0)$ for the $^{27}$Al$^+$ ion. Other theoretical and experimental results are also presented for comparison. Numbers in square brackets stand for the power of $10$ and in parentheses for the uncertainties.\label{data-Al}}
\begin{ruledtabular}
\begin{tabular}{lcccccccccc}
& \multicolumn{2}{c}{Zeeman} && \multicolumn{2}{c}{Hyperfine} && \multicolumn{2}{c}{$\Delta E$ } && \\
\cline{2-3}\cline{5-6}\cline{8-9}
Model & $\langle ^3\!P^o_0 ||- {\bm \mu}^{(1)}|| ^3\!P^o_1 \rangle$ & $\langle ^3\!P^o_0 ||-{\bm \mu}^{(1)}|| ^1\!P^o_1 \rangle$ && $\langle ^3\!P^o_1 ||\bm{T}^{(1)}|| ^3\!P^o_0 \rangle$ & $\langle ^1\!P^o_1 ||\bm{T}^{(1)}|| ^3\!P^o_0 \rangle$ && $^{3}\!P^o_1~-~^{3}\!P^o_0$ & $^{1}\!P^o_1~-~^{3}\!P^o_0$ && $\delta g^{(1)}_{\rm hfs}(^3\!P^o_0)$ \\
\hline
DHF	      & $-$7.0873[$-$1] & 1.62[$-$3] && 6.037[$-$2] & 4.591[$-$2] && 2.914[$-$4] & 1.485[$-$1] && $-$9.814[$-$3]  \\ 
CV        & $-$7.0872[$-$1] & 2.56[$-$3] && 7.077[$-$2] & 5.602[$-$2] && 2.989[$-$4] & 1.038[$-$1] && $-$1.122[$-$3]  \\ 
CC$_{2}$  & $-$7.0872[$-$1] & 2.34[$-$3] && 6.960[$-$2] & 5.489[$-$2] && 3.000[$-$4] & 1.141[$-$1] && $-$1.099[$-$3]  \\ 
MR 	      & $-$7.0873[$-$1] & 2.46[$-$3] && 6.932[$-$2] & 5.443[$-$2] && 2.983[$-$4] & 1.060[$-$1] && $-$1.101[$-$3]  \\ 
Breit     & $-$7.0873[$-$1] & 2.33[$-$3] && 6.936[$-$2] & 5.448[$-$2] && 2.777[$-$4] & 1.060[$-$1] && $-$1.183(6)[$-$3]
           \\
\multicolumn{11}{c}{Theories} \\
Itano~\textit{et al.}~\cite{Itano2007}          & & && 6.931[$-$2] & 5.487[$-$2] && & && $-$1.181[$-$3] \\
Andersson~\textit{et al.}~\cite{Andersson2010}  & & && 7.084[$-$2] & 5.623[$-$2] && & && \\
Kang~\textit{et al.}~\cite{Kang2009}            & & && 6.932[$-$2] & 5.537[$-$2] && & && \\
Beloy\textit{et al.}~\cite{Beloy2017}           & & && 6.899[$-$2] &         && & && \\
Safronova~\textit{et al.}~\cite{Safronova2000}  & & &&         &         && 2.802[$-$4] &  &&  \\
J\"onsson~\textit{et al.}~\cite{Jonsson1997}    & & &&         &         && 2.757[$-$4] &   &&   \\
Zou~\textit{et al.}~\cite{Zou2001}              & & &&         &         && 2.799[$-$4] & 1.036[$-$1] &&            \\
\multicolumn{11}{c}{Experiments} \\
NIST~\cite{NIST}                                & & &&         &         && 2.774[$-$4] & 1.023[$-$1] &&              \\
Rosenband~\textit {et al.}~\cite{Rosenban2007}  & & &&         &         &&  &  && $-$1.18437(8)[$-$3]\\
\end{tabular}
\end{ruledtabular}
\end{table}
\end{turnpage}
\subsection{The case of $^{87}$Sr}
In the case of Sr, we also started from the DHF calculation to optimize the spectroscopic orbitals occupied in reference configuration $1s^22s^22p^63s^23p^63d^{10}4s^24p^65s5p$. $5s$ and $5p$ were regarded as the valence orbitals and others as the core. In the following SCF calculations, the VV and major CV electron correlations were taken into account. The major CV electron correlation includes those between electrons in the valence and $n \ge 3$ core shells. The virtual orbitals were added layer by layer up to $n=11$ and $l=4$. 
 
Keeping all orbitals frozen, we further considered the effect of the CV correlation related to the $n \le 2$ electrons in the subsequent RCI computation. This model is labelled as CV. The CC electron correlation in the $4s$ and $4p$ subshells, referred to as CC$_{4}$, was also captured in RCI. To control the number of CSFs, only the first five layers of virtual orbitals were used to generate the CSFs accounting for the CC correlation. Higher-order correlation among $n \ge 4$ electrons was considered by the MR-SD approach. The MR configurations are composed of \{$4s^24p^65s5p$, $4s^24p^64d5p$, $4s^24p^65s6p$, $4s^24p^65p6s$, $4s^24p^64d6p$\}. The corresponding configuration space was expanded by SD-excitation CSFs from the MR configuration set to the first five layers of the virtual orbitals. Finally, the Breit interaction was evaluated based on the MR model.

In Table \ref{g-Sr} we display $g$-factors of the $5s5p~^{3,1}\!P^o_1$ states obtained with various computational models for Sr. It is worth noting that the contribution from the CC$_4$ and higher-order correlations to the $g$-factors is comparable with those from the VV and CV correlations. As found in the case of Al$^+$, the effect of the higher-order electron correlation on the $g$-factors compensates to that of the CC correlation. Therefore, both of them should be included. The QED corrections, about $10^{-9}$ to the $g(^{3,1}\!P^o_1)$, are fractional, and thus omitted in our calculations. The computational uncertainties mainly arise from the neglected CC and higher-order electron correlations related to the $n \le 3$ core shells. Nevertheless, these effects should be smaller than those from the outer shells because of the stronger nuclear Coulomb potential in the inner region. Additionally, with respect to the cancellation between the CC and MR electron correlations in the $n=4,5$ subshells, we estimated the uncertainties to be about $4 \times 10^{-5}$ for the $g$-factor of the $^3\!P^o_1$ state and about $7 \times 10^{-5}$ for the $^1\!P^o_1$ state. The present results are consistent with the measurement for $^3\!P^o_1$~\cite{Ma1968}.

\begin{table}
\caption{$g$-factors of the $5s5p~^{3,1}\!P^o_1$ states for Sr. The number in parentheses stands for the uncertainties. \label{g-Sr}}
\begin{ruledtabular}
\begin{tabular}{ldldl}      
Model                   &&    $g(^3\!P^o_1)$  &&     $g(^1\!P^o_1)$   \\\hline
DHF                     &&      1.501014      &&      1.000104        \\
CV                      &&     1.500624       &&      1.000508        \\
CC$_{4}$                &&     1.500888       &&      1.000237        \\
MR-SD                   &&     1.500658       &&      1.000439        \\
Breit                   &&     1.500673(40)   &&      1.000425(72)    \\
NR                      &&     1.5            &&      1.0             \\
Ma~\cite{Ma1968}        &&     1.50065(4)     &&                      \\
\end{tabular}
\end{ruledtabular}
\end{table}

Matrix elements of the Zeeman and hyperfine interactions, $\langle^3\!P^o_0 ||-{\bm \mu}^{(1)}|| ^{3,1}\!P^o_1\rangle$ and $\langle^{3,1}\!P^o_1 \|\bm {T}^{(1)}\|^3\!P^o_0\rangle$, energy separations $\Delta E(^{3,1}\!P^o_1$ - $^3\!P^o_0)$ and hyperfine-induced Land\'e $g$-factor $\delta g^{(1)}_{\rm hfs}(^3\!P^o_0)$ are presented in Table~\ref{data-Sr} as functions of the computational models for the $^{87}$Sr atom. It was found that the dominant corrections are made by the VV and CV electron correlations to all physical quantities concerned. As emphasized in the case of Al$^+$, the effects of the CC and higher-order electron correlations and the Breit interaction are also indispensable to achieve satisfactory accuracy for Sr. It is more difficult to estimate the computational error for the case of Sr, because of the extra $n=3$ shell compared with Al$^+$. According to our test, the contribution from the $n=3$ CC correlation reaches 4\% to $\delta g^{(1)}_{\rm hfs}(^3\!P^o_0)$. Hence, we gave a rough uncertainty of $3 \times 10^{-6}$ to the final result. Our calculated hyperfine-induced $g$-factor of the $^3\!P^o_0$ clock state is in good agreement with recent experimental values for $^{87}$Sr~\cite{Boyd2007, Shi2015}.

\begin{turnpage}
\begin{table}
\caption{Matrix elements (in a.u.) of Zeeman and hyperfine interactions, $\langle^3\!P^o_0 ||-{\bm \mu}^{(1)}||^{3,1}\!P^o_1\rangle$ and $\langle^{3,1}\!P^o_1 \|\bm {T}^{(1)}\|^3\!P^o_0\rangle$, energy intervals (in a.u.) $\Delta E(^{3,1}\!P^o_1~-~^{3}\!P^o_0)$, and $\delta g^{(1)}_{\rm hfs}(^3\!P^o_0)$ for the $^{87}$Sr atom. Other theoretical and experimental results are also presented for comparison. Numbers in square brackets stand for the power of $10$ and in parentheses for the uncertainties. \label{data-Sr}}
\begin{ruledtabular}
\begin{tabular}{lcccccccccc}

 & \multicolumn{2}{c}{Zeeman} && \multicolumn{2}{c}{Hyperfine } && \multicolumn{2}{c}{$\Delta E$ } &&  \\
\cline{2-3}\cline{5-6}\cline{8-9}
Model & $\langle ^3\!P^o_0 ||-{\bm \mu}^{(1)}|| ^3\!P^o_1\rangle$ & $\langle ^3\!P^o_0 ||-{\bm \mu}^{(1)}||^1\!P^o_1 \rangle$ && $\langle^3\!P^o_1 ||\bm{T}^{(1)}||^3\!P^o_0\rangle$ & $\langle^1\!P^o_1 ||\bm{T}^{(1)}||^3\!P^o_0\rangle$ && $^{3}\!P^o_1$ - $^{3}\!P^o_0$ & $^{1}\!P^o_1$ - $^{3}\!P^o_0$ && $\delta g^{(1)}_{\rm hfs}(^3\!P^o_0)$ \\
\hline 
DHF         & $-$7.0863[$-$1] & 7.087[$-$3]  && 6.981[$-$2] & 5.068[$-$2] && 8.323[$-$4] & 6.976[$-$2] && 6.63[$-$5]     \\ 
CV          & $-$7.0834[$-$1] & 2.054[$-$2]  && 9.351[$-$2] & 6.806[$-$2] && 8.729[$-$4] & 3.371[$-$2] && 8.46[$-$5]     \\ 
CC$_{4}$    & $-$7.0855[$-$1] & 1.317[$-$2]  && 8.517[$-$2] & 6.373[$-$2] && 8.806[$-$4] & 4.942[$-$2] && 7.64[$-$5]     \\ 
MR          & $-$7.0837[$-$1] & 1.962[$-$2]  && 8.341[$-$2] & 5.991[$-$2] && 8.640[$-$4] & 3.506[$-$2] && 7.62[$-$5]     \\ 
Breit       & $-$7.0839[$-$1] & 1.930[$-$2]  && 8.372[$-$2] & 6.016[$-$2] && 8.496[$-$4] & 3.499[$-$2] && 7.78(30)[$-$5] \\ 
\multicolumn{11}{c}{Theories} \\
Boyd \textit{et al.}~\cite{Boyd2007}          &&&&&&&&&& 7.795(7)[$-$5]\footnotemark[1]   \\
                                              &&&&&&&&&& 8.42(4)[$-$5]\footnotemark[2]   \\
         \multicolumn{11}{c}{Experiments} \\
NIST~\cite{NIST}                              &&&&&&& 8.512[$-$4] & 3.363[$-$2] &&          \\
Takamoto~\cite{Takamoto2003}                  &&&&&&&&&& 7.573[$-$5]\footnotemark[3]        \\
Boyd {\it {et al.}}~\cite{Boyd2007}           &&&&&&&&&& 7.74(3)[$-$5]\footnotemark[3]     \\
Shi \textit{et al.}~\cite{Shi2015}            &&&&&&&&&& 7.746(5)[$-$5]\footnotemark[3]     \\
\end{tabular}
\footnotetext[1]{Calculated by BW method.}
\footnotetext[2]{Calculated by MBW method.}
\footnotetext[3]{In practice, the differential $g$-factors of the $5s^2~^1\!S_0~-~5s5p~^3\!P^o_0$ clock transition were measured for $^{87}$Sr, but this is equivalent to the hyperfine-induced Land\'e $g$-factor of $5s5p~^3\!P^o_0$ (See Sec.~\ref{g}).}
\end{ruledtabular}
\end{table}
\end{turnpage}

\section{Conclusion}
In the weak-magnetic-field approximation, the expression was given for hyperfine-induced Land\'{e} $g$-factors. Using the MCDHF method, we calculated $\delta g^{(1)}_{\rm hfs}(^3\!P^o_0)$ for the $^{27}$Al$^+$ ion and the $^{87}$Sr atom. Also, we investigated the effects of electron correlations and relativity on the atomic parameters involved in $\delta g^{(1)}_{\rm hfs}(^3\!P^o_0)$. The present results are in good agreement with other measurements, especially for the case of $^{27}$Al$^{+}$ in which the computational uncertainty was controlled at the level of $6 \times 10^{-6}$. This indicates that our theory can be generalized to predict hyperfine-induced Land\'e $g$-factors for other atomic systems.

\begin{acknowledgments}
This work was supported by the Nation Natural Science Foundation of China under Grant No. 11874090 and No. 11934014, the Strategic Priority Research Program of the Chinese Academy of Sciences under Grant No. XDB21030300, the National Key Research and Development Program of China under Grant No. 2017YFA0304402, and the West Light Foundation of the Chinese Academy of Sciences under Grant No. XAB2018B17.
\end{acknowledgments}

\bibliography{ttlib}
\end{document}